%% file: paper-submit.tex
\documentclass[aps,pra,showpacs]{revtex4}
\usepackage{bbm}
\usepackage{mathrsfs}
\usepackage{psfrag,epsfig}
\usepackage{amsmath,amsfonts,amssymb}

\include{makros2}

\begin{document}

\preprint{draft}

\title{Towards a Theory of Molecular Forces between Deformed Media}

\author{Rauno B\"uscher and Thorsten Emig}

\affiliation{Institut f\"ur Theoretische Physik, Universit\"at zu
K\"oln, Z\"ulpicher Stra\ss e 77, D-50937 K\"oln, Germany}

\date{\today}

\begin{abstract}
  A macroscopic theory for the molecular or Casimir interaction of
  dielectric materials with arbitrarily shaped surfaces is developed.
  The interaction is generated by the quantum and thermal fluctuations
  of the electromagnetic field which depend on the dielectric function
  of the materials. Using a path integral approach for the
  electromagnetic gauge field, we derive an effective Gaussian action
  which can be used to compute the force between the objects. No
  assumptions about the independence of the shape and material
  dependent contributions to the interaction are made.  In the
  limiting case of flat surfaces our approach yields a simple and
  compact derivation of the Lifshitz theory for molecular forces
  \cite{Lifshitz}.  For ideal metals with arbitrarily deformed
  surfaces the effective action can be calculated explicitly. For the
  general case of deformed dielectric materials the applicability of
  perturbation theory and numerical techniques to the evaluation of
  the force from the effective action is discussed.
\end{abstract}

\pacs{12.20.-m, 33.90.+h, 42.50.Ct}

\maketitle

\section{Introduction}

The last years have witnessed a resurgence of theoretical and
experimental research on Casimir interactions between macroscopic and
mesoscopic objects
\cite{Bordag+01,Milton01,milonni,Kardar+99,Mostepanenko+97,Israelachvili92}.
The most mentionable recent achievements include on the experimental
side the high precision measurements of the attractive Casimir force
between metallic surfaces
\cite{Lamoreaux97,Lamoreaux98,Mohideen+98,Lambrecht+00,Lamoreaux00,Ederth00,Chan+01,Chan+01b,Bressi+02,Decca+03}
and the simultaneous study of both the critical Casimir force due to
(thermal) order parameter fluctuations and the electrodynamic Casimir
force in superfluid Helium films near the critical point
\cite{Garcia+99,Garcia+02,Ueno+03}. On the theoretical side there is
an ongoing attempt to describe the Casimir interaction between
non-ideal metals with finite conductivity, including simultaneously
the effect of finite temperature, see \cite{Geyer+03} for a recent
summary of the used and partly controversial approaches.  Another
direction of great interest is the study of the strong geometry
dependence of the Casimir force which is inevitably linked to the
non-additivity of fluctuation induced forces
\cite{Emig+01,Emig+03,emig,Genet+03}. Even a potential change from
attractive to repulsive forces due to either the material properties
\cite{Kenneth+02,Iannuzzi+03,Kenneth+03} or the geometry of the
objects \cite{Maclay00} has been discussed. As to the comparison
between experiment and theory, Casimir's prediction \cite{Casimir48}
\begin{equation}
  \label{eq:original-force}
  \frac{F_0}{A}=-\frac{\pi^2}{240} \frac{\hbar c}{H^4}
\end{equation}
for the force $F_0$ between ideal metallic and parallel flat surfaces
of area $A$ and distance $H$ at zero temperature has been confirmed by
the recent experiments within a few percent of accuracy.  However,
there is broad agreement that the experiments have shown that the
inclusion of the material properties of the surfaces, of the roughness
and geometry of the surfaces, and of finite temperature effects is
indispensable. A solid theoretical account of these effects is
necessary in view of the importance of the experimental results to the
test of unified gauge theories of fundamental interactions
\cite{Fischbach+01} and the design of nanotechnological devices
\cite{Chan+01,Buks+01}. 

Each of the modifications to the ideal prediction of
Eq.~(\ref{eq:original-force}) introduces at least one corresponding
length scale in the interaction energy. If the interacting materials
are not ideal metals, the electromagnetic fluctuations are not
reflected perfectly at all wavelengths by the surfaces. Then one
expects Eq.~(\ref{eq:original-force}) to hold only for distances $H$
much larger than the plasma wavelength $\lambda_p$ of the metal. In
the recent experiments $\lambda_p$ is of the order of $0.1\mu$m. At
smaller distances the force will be reduced to $F=\eta_\text{m} F_0$
by a material dependent factor $\eta_\text{m}<1$ compared to the ideal
force. Since the experiments are usually performed at room
temperature, thermal fluctuations of the electromagnetic field tend to
increase the force for separations $H$ which are larger than the de
Broglie wavelength of photons, $\lambda_\text{T}=\hbar c/(k_\text{B}
T)$ ($\approx 7\mu$m at $300^\circ$K), leading to $F=\eta_\text{T}
F_0$ with the temperature dependent factor $\eta_T>1$. Finally the
presumably most difficult to calculate change of the force $F_0$ comes
from the surface geometry. The geometry modifications can be divided
into two different types. The first is a in general unwanted
stochastic surface roughness, the second an intentionally designed
surface structure like a corrugation. The latter has been studied in a
recent experiment to study the geometry dependence of the normal and
lateral Casimir force \cite{Roy+99}. Both types can be characterized
by the deformation amplitude $a$ and the roughness correlation or
corrugation length $\lambda$. Recent theoretical work for ideal metals
at zero temperature has shown that the force is generally increased by
deviations from the flat surface geometry, at least for uni-axial
structures \cite{Emig+01,Emig+03,emig}. Therefore we can write
$F=\eta_\text{g} F_0$ with a factor $\eta_\text{g}$ accounting for
geometry dependent changes of the force.

So far, we have discussed the different modifications of the ideal
force $F_0$ independently. One can expect that this is indeed
justified if the characteristic length scales of the modifications are
widely different from each other. However, this is by no means always
the case in the recent experiments, especially at the point of closest
approach of the interacting surfaces, and in nanotechnological
devices. In view of this, it is of importance to consider correlations
between the force modifications. That means that one has to assume a
more general form of the force
\begin{equation}
  \label{eq:force-corrected}
  F=\eta_\text{m}\, \eta_\text{T}\, \eta_\text{g}\, 
( 1 + \Delta_\text{corr} ) F_0
\end{equation}
with a new term $\Delta_\text{corr}$ accounting for correlations
between different modifications. The present state of the techniques
available in literature, however, does not allow to determine
$\Delta_\text{corr}$ in general. A commonly used but uncontrolled
approximation is to set $\Delta_\text{corr}=0$. Most of the recent
theoretical studies of correlation effects have been devoted to the
simultaneous effect of finite conductivity of metallic surfaces and
finite temperature, see \cite{Geyer+03} for a recent overview and
\cite{Chen+03} for a proposed experiment to measure these effects.
Common to almost all of the existing theoretical work on the
correlations between conductivity and temperature corrections is that
it starts from the so-called Lifshitz theory for molecular
interactions between macroscopic objects \cite{Lifshitz}. The Lifshitz
theory provides a formula for the force in the rather general case of
dielectric bodies at arbitrary temperature but with flat and parallel
surfaces.  The key difference in the approaches to correlation effects
is the treatment of the zero Matsubara frequency term of the Lifshitz
formula when it is applied to different models (Drude, ideal metal or
free plasma model) for the dielectric function of the metals. In
addition, there are a few approaches which are not based on the
Lifshitz theory but employ a surface impedance boundary condition in
order to account for the coupling between electromagnetic fluctuations
and a metallic surface. To date, however, it appears that there is no
broad agreement on a correct prescription for the evaluation of the
Lifshitz formula in the case of non-ideal metallic surfaces at finite
temperatures.

The situation becomes even worse when one considers deviations from
the parallel flat plate geometry and correlations between geometry,
conductivity and temperature modifications of the Casimir force. To
our knowledge there is no complete theory for correlations between
geometry induced modifications of the force and the above discussed
corrections available in literature. In fact, only recently an exact
description for the sole geometry dependence of the force, making no
additivity assumptions of two-body forces, has been given
\cite{Emig+01,Emig+03,emig}.  So far, surface roughness in combination
with finite conductivity has been studied only by neglecting
correlations between both effects \cite{Klimchitskaya+99}. However,
the characteristic length scales of surface roughness or designed
surface corrugations in nanotechnological devices can be close to the
relevant length scales of the material as, e.g., the plasma
wavelength. Therefore it would be very useful to have an analog of the
Lifshitz theory for more general geometries.  Such a general theory
should yield the Casimir interaction depending on the dielectric
function of the material, temperature and a height profile describing
the surface geometry as input parameters. The development of such a
theory is the purpose of the present work.

In this paper we will introduce a novel macroscopic approach to
molecular forces between dielectric media. It will be based on a path
integral technique for fluctuation induced forces which was previously
developed for ideal metals at zero temperature
\cite{likar,Kardar+99,Emig+01}.  Non-local boundary conditions for the
electromagnetic gauge field are employed to treat the interaction
between electromagnetic fluctuations and matter. The boundary
conditions can be viewed as a reformulation of the so-called
extinction theorem of classical electrodynamics \cite{ewald,oseen,bw}.
The important new property of our approach is that surface
deformations can be included without any assumption about the
correlations between contributions to the force from geometrical and
material properties.  We derive an effective Gaussian action which is
a functional of the frequency dependent dielectric function of the
material and the height profile of the surfaces. The effective action
is a possible starting point for future detailed analyses of the
effect of correlation on the force as described by the
$\Delta_\text{corr}$ in Eq.~(\ref{eq:force-corrected}). We demonstrate
the efficiency of our approach by looking at two particular limits of
interest. First, we consider flat surfaces of dielectric media. In
this case we obtain, as a byproduct of our theory, a compact and
concise derivation of the Lifshitz formula for molecular forces
\cite{Lifshitz} in the language of quantum statistical mechanics. An
even simpler derivation of this formula is found by a scalar field
approach, cf.~the Appendix, which should be compared to other
derivations \cite{VanKampen+68,Langbein72} of the original result of
Lifshitz. As second limiting case, we consider ideal metals with
arbitrary deformations. Then the effective action assumes a simple
form which can be determined explicitly.  For general deformed
dielectric media, the effective action can be used as a basis for
perturbative \cite{Emig+01,Emig+03} or numerical \cite{emig}
computations of the correlation term $\Delta_\text{corr}$.

The outline of the rest of the paper is as follows. In the following
section we develop the path integral approach for deformed surfaces of
materials which are characterized by a general frequency dependent
dielectric function. We derive the non-local boundary conditions which
describe the reflection and refraction properties of the interacting
bodies.  By integrating out the electromagnetic gauge field, an
effective action for the interaction between the bodies is obtained.
In Sec.  \ref{sec:lt} we apply our theory to calculate the force
between two flat surfaces of dielectric media. In this case, the known
Lifshitz formula for molecular forces is found.  The effective action
for deformed surfaces of ideal metals is computed explicitly in
section \ref{sec:ideal-metal}. Section \ref{sec:discussion} provides a
discussion of the relevance of our results to perturbative and
numerical analyses of the Casimir or molecular interaction between
macroscopic objects. A rather short and concise derivation of the
standard Lifshitz theory in terms of a scalar field is left to the
Appendix.

\section{Path-integral formulation of molecular forces}
\label{sec:pi}

We will develop a macroscopic theory which allows to calculate the
interaction between materials of rather general shape. Instead of
considering directly the field emitted by the fluctuating dipoles in
the material we view the interaction as occurring through the
modifications of the quantum (and thermal) fluctuations of the
electromagnetic field {\it between} the materials. No direct reference
is made to the electromagnetic field fluctuations in the interior of
the materials. The effect of the dipoles induced by the external
fluctuating field will be described by material dependent boundary
conditions which are defined on the surface of the material. Our
method is based on a path integral quantization of the electromagnetic
gauge field. This approach has full generality in the sense that it
can be applied to any body, characterized by its dielectric function,
with any surface profile, described by a height field, at any
temperature. 

The common approaches for computing the force between materials is to
first determine the solution of Maxwell's equations both inside and
outside the materials, and than to evaluate the force either from the
stress tensor or from the zero-point energy of the modes using the
so-called argument theorem, see, e.g., Ref. \cite{milonni}. The
problem with these approaches is that they are not suited to treat
non-flat surfaces since deformations in general lead to a complicated
modification of the mode structure and make the solution to Maxwell's
equations a hard task. In the following, we will formulate the
interaction between deformed materials within the language of quantum
statistical mechanics. Since this formulation makes no explicit use of
the individual eigenfrequencies of the modes it is better targeted for
the treatment of deformations. 
 
\begin{figure}[b]
\includegraphics[width=0.7\textwidth]{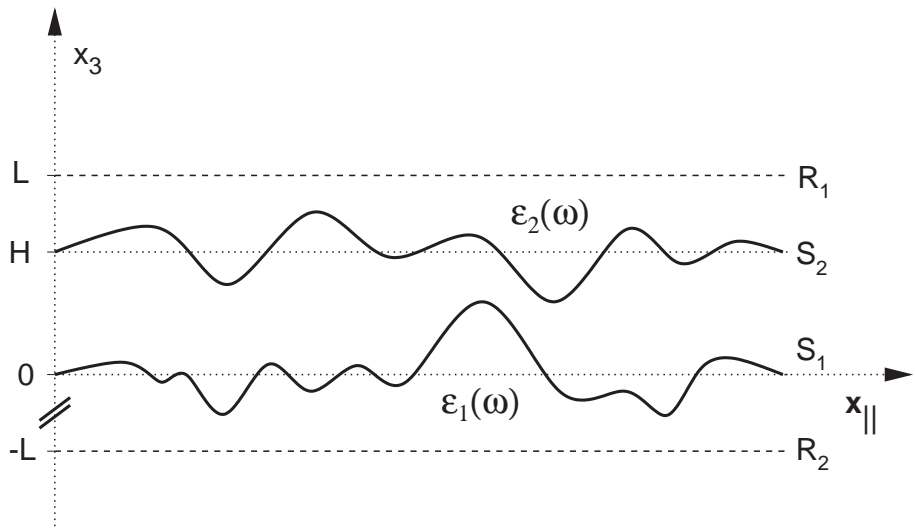}
\caption{\label{fig:surface} 
  Two deformed surfaces $S_1$ and $S_2$ of dielectric media with
  dielectric functions $\epsilon_1(\omega)$ and $\epsilon_2(\omega)$,
  respectively, separated by a gap of mean size $H$ along the
  $x_3$--direction. The meaning of the auxiliary surfaces $R_1$ and
  $R_2$ is explained in the text.}
\end{figure}

We consider the two interacting media as filling half spaces which are
bounded by deformed surfaces $S_\alpha$, $\alpha=1,2$. The
deformations from a plan-parallel geometry of mean surface distance
$H$ are described by the height functions $h_\alpha(\xbf_\|)$ with
$\xbf_\|$ the lateral surface coordinates, see Fig.~\ref{fig:surface}.
The media are characterized by their complex dielectric functions
$\epsilon_\alpha(\omega)$, respectively. The gap between the media is
assumed to be vacuum, $\epsilon(\omega)=1$.  The free energy $\Fcal$ of
the photon gas in the gap between the two surfaces can be calculated
from the imaginary time path integral for the electromagnetic gauge
field $A$. In the absence of media, the vacuum partition function
$\Zcal_0$ is given by
\begin{equation}
\Zcal_0^2\:=\:\int\!
\Dsc\big(A^\ast A\big)\:e^{-S_{\text E}(A^\ast\!\!,A)},
\end{equation}
where we have introduced a complex valued gauge field which leads to a
double counting of each degree of freedom. The reason for this will
become clear below when we discuss the boundary conditions at the
surfaces. The Euclidean action $S_{\text E}(A^\ast\!\!,A)$ is the
imaginary time version of the action $S(A^\ast\!\!,A)$ of the
electromagnetic field in the Minkowski space time with coordinates
$\:X=(t,\xbf)=(t,\xbf_\|,x_3)\:$, 
\begin{equation}\label{eq:parameter_action_0}
S(A^{\ast}\!\!,A)\:=\:-\frac{1}{2}
\int_{X}\!\big(F^\ast_{\mu\nu}F^{\mu\nu}\big)(X)-\frac{1}{\xi}
\int_{X}\!\big(\pd_\mu A^{\ast\mu}\big)\big(\pd_\nu A^{\nu}\big)(X)
\end{equation}
where the first term comes from the Lagrangian of the electromagnetic
field $\:F^{\mu\nu}=\pd^\mu A^\nu-\,\pd^\nu A^\mu\:$ and the second
term results from the Fadeev-Popov gauge fixing procedure which
assures that each physical field configuration is counted only once in
the path integral over the gauge field. The parameter $\xi$ allows to
switch between different gauges; all gauge invariant quantities
calculated from this action like, e.g., the Casimir force, do not
depend on $\xi$. In the following we will use the Feynman gauge
corresponding to $\xi=1$. The coefficients in the action of
Eq.~(\ref{eq:parameter_action_0}) differ by a factor of $1/2$ from the
conventional definition of the action for a real valued gauge field in
order to obtain the correct photon propagator which in Feynman gauge
reads $G_{\mu\nu}=g_{\mu\nu}/K^2$ with momentum $\:K=(\omega,\kbf)\:$,
$\:K^2=K_\mu K^\mu=\omega^2-\kbf^2\:$ and Minkowski metric tensor
$g_{\mu\nu}=\text{diag}(1,-1,-1,-1)$. The Euclidean action
$S_\text{E}(A^\ast,A)$ is obtained from
Eq.~(\ref{eq:parameter_action_0}) by applying a Wick rotation to
imaginary time which amounts to the transformations
$
t \to -i\tau, \quad \omega \to i \zeta$ and 
$A^0 \to iA^0, \quad
A^{\ast 0}\to iA^{\ast 0},
$
while the remaining components $A^\mu$ remain unchanged
\cite{Nieuwenhuizen+96a,Nieuwenhuizen+96b}.  Since this transformation
corresponds to the change $g_{\mu\nu} \to -\delta_{\mu\nu}$ for the
metric tensor, the Euclidean action in momentum space becomes
\begin{equation}
\label{eq:action-e}
S_E(A^{\ast}\!,A)\:=\:
\frac{1}{\beta} \sum_{n=-\infty}^\infty \int_{\kbf}\!A^{\ast\mu}(\zeta_n,\kbf)
\Gcal^{-1}_{\text{E},\mu\nu}(\zeta_n,\kbf)
A^\nu(\zeta_n,\kbf)
\end{equation}
where we allowed for a finite temperature $T$ by introducing bosonic
Matsubara frequencies $\zeta_n=2\pi n/\beta$ with $\beta=1/T$. The
Euclidean Green function is given by $\Gcal_{\text{E},
  \mu\nu}(\zeta,\kbf)=\delta_{\mu\nu}\Gcal(\zeta,\kbf)$ with
$\Gcal(\zeta,\kbf)=(\zeta^2+\kbf^2)^{-1}$. 

In the presence of the two media the free energy of the photon gas in
the gap between the media is obtained from a restricted partition
function. The restrictions are due to the boundary conditions for the
gauge field which are imposed by the dielectric properties of the
media. It will be shown below that there are three boundary conditions
on each surface $S_\alpha$ which we number by $j=1,2,3$. Each of these
conditions implies the vanishing of a {\it non-local} linear
combination of derivatives of the components of the gauge field. Thus
the mean distance $H$-dependent restricted partition function
$\Zcal(H)$ can be written as
\begin{equation}\label{eq:NPF}
\Zcal^2(H)\:=\:\frac{1}{\Zcal_0^2}\int\!
\Dsc\left(A^\ast A\right)
\:\prod_{\alpha,j} \prod_{\zeta_n}\prod_{\xbf \in R_\alpha} 
\delta\left[\int_{\xbf'\in S_\alpha} \Lcal_{j\mu}^\alpha(\zeta_n;\xbf,\xbf')
A^\mu(\zeta_n,\xbf')  \right]
\:e^{-S_\text{E}(A^\ast\!\!,A)},
\end{equation}
where we enforced the boundary conditions by inserting delta-functions
for all positions $\xbf$ on (flat) auxiliary surfaces $R_\alpha$ which
are placed at $x_3=\pm L$ with sufficiently large $L$ so that the
surfaces $S_\alpha$ are located between them, see
Fig.~\ref{fig:surface}. The final result for the force between the
media will be of course independent of $L$.  The differential
operators $\Lcal_{j\mu}^\alpha$ depend via both the dielectric
function $\epsilon_\alpha$ and the normal vector $\hat \nbf_\alpha$ on
the surface index $\alpha$. Their actual form will be computed below.
The interaction (Casimir) free energy per unit area of the two
surfaces $S_\alpha$ is then given by
\begin{equation}
\label{def-energy}
\Fcal(H)=-\frac{1}{A\beta} \ln \left[ \Zcal(H) 
\Zcal^{-1}_\infty(H)\right], 
\end{equation}
where $A$ is the surface area and $\beta=1/T$ the inverse temperature.
$\Zcal_\infty$ is the asymptotic limit of $\Zcal$ for large $H$ so
that $\Fcal$ is measured relative to two surfaces which are infinitely
apart from each other. The force per unit area between the surfaces is
then given by $F=-\partial \Fcal/\partial H$.

\subsection{Boundary conditions}
\label{sec:bcs}

In this section we will derive the boundary conditions at the surfaces
of the dielectric media. The boundary conditions are based on the
optical extinction theorem of Ewald \cite{ewald} and Oseen
\cite{oseen}, see also \cite{bw}.  This theorem states that part of
the electromagnetic field produced by the molecular dipoles inside a
medium exactly cancels the incident field, while the remainder
propagates according to Maxwell's equations in continuous media.
Ewald and Oseen proved the theorem for crystalline media and
amorphous, isotropic dielectrics, respectively, from a point of view
of classical molecular optics.  Later, Born and Wolf extended the
theorem to more general classes of materials \cite{bw}. A relationship
between the extinction theorem and the Lifshitz theory of dispersion
forces for continuous media has been pointed out by Milonni and Lerner
\cite{Milonni+92}. They use the fact that the extinction theorem
permits a reduction of the multiple-scattering problem for the
molecular dipoles to the solution of the wave equation for the gauge
field $A$ with appropriate boundary conditions. From this they
conclude that the extinction theorem shows that the macroscopic
Lifshitz theory for continuous media correctly accounts for all
multiple-scattering {\it nonadditive} contributions to the force
between flat surfaces. We will demonstrate that these concepts are
useful to describe the interaction of even deformed surfaces.

In the following we use an (equivalent) reformulation of the
extinction theorem as a non-local boundary condition which enforces
the laws of reflection and refraction at the surfaces of the
interacting media. Our derivation of the boundary conditions follows
closely the approach outlined in \cite{marvin}. Let us start with the
Helmholtz wave equation for the magnetic field $\Bbf$ inside a
dielectric medium which occupies the volume $V$ with surface
$S=\partial V$,
\begin{equation}\label{eq:helmholtz}
\Big[\nabla^2+\epsilon(\omega)\omega^2\Big]\Bbf(\omega,\xbf)\:=\:0.
\end{equation}
The propagation of the field inside the medium is described by the
material Green function satisfying
\begin{equation}\label{eq:greenfunction}
\Big[\nabla'^2+\epsilon(\omega)\omega^2\Big]\Gcal^\epsilon
(\omega;\xbf,\xbf')\:=\:\delta^{(3)}\big(\xbf-\xbf'\big),
\end{equation}
where we allow for a general frequency dependent complex dielectric
function $\epsilon(\omega)$. Applying Green's theorem to the
components of $\Bbf$ and to $\Gcal^\epsilon$, one easily obtains,
using (\ref{eq:helmholtz}) and (\ref{eq:greenfunction}),
\begin{equation}\label{eq:greens_theorem}
\int_{\xbf' \in S}\!\Big[
\Gcal^\epsilon
(\omega;\xbf,\xbf')\big(\hat{\nbf}'\cdot\nabla'\big)\Bbf(\omega,\xbf')
\:-\:
\Bbf(\omega,\xbf')\big(\hat{\nbf}'\cdot\nabla'\Gcal^\epsilon
(\omega;\xbf,\xbf')\big)
\Big]\:=\: \left\{
\begin{array}{lcl}
\Bbf(\omega,\xbf) & , & \xbf \in V\\
0 & , & \xbf \notin V 
\end{array}
\right.
\end{equation}
with the normal unit vector $\hat\nbf$ of the surface pointing into the
vacuum. We will make use of this result for the case where the
position $\xbf$ is located on an auxiliary surface which is placed
outside the medium so that the integral has to vanish. Using a
succession of vector identities \cite{jackson} and the macroscopic
Maxwell equations for continuous media, $\nabla \cdot \Bbf =0$,
$\nabla \times \Bbf = -i\omega \epsilon(\omega) \Ebf$, the condition
that the integral vanishes can be transformed to
\begin{equation}\label{eq:EX2}
\int_{\xbf' \in S}
\Big[\!-\!i\omega\epsilon(\omega)\big(\hat\nbf'
\times\Ebf(\omega,\xbf') \big)\:+\:
\big(\hat\nbf' \cdot\Bbf(\omega,\xbf')\big)\nabla'\:+\:
\big(\hat\nbf' \times\Bbf(\omega,\xbf')\big)\times\nabla'
\Big]\Gcal^\epsilon(\omega;\xbf,\xbf')
=0.
\end{equation}
The usual continuity conditions for the electric and magnetic field at
dielectric surfaces \cite{jackson} without external surface charges
and currents show that the three terms in Eq.~(\ref{eq:EX2}) have to
be continuous across the surface. Therefore we can use the vanishing
of the integral as a boundary condition for the electromagnetic field
on the vacuum side of the surface. As a side remark we note that if we
had started with the wave equation for the electric field instead of
the magnetic field we had obtained a similar expression as that in
Eq.~(\ref{eq:EX2}). However, due the discontinuity of the dielectric
function across the surface the condition that the integral vanishes
had not translated to the field on the vacuum side. Finally, in the
case of an ideal metal, $\epsilon(\omega) \to \infty$, and the
integral in Eq.~(\ref{eq:EX2}) is dominated by the first term. In this
limit the integration can be carried out, leading to the well-known
condition $\hat \nbf \times \Ebf=0$ for ideal metals.

The condition of Eq.~(\ref{eq:EX2}) can now be used to determine the
differential operators $\Lcal^\alpha_{j \mu}(\zeta,\xbf,\xbf')$
appearing in Eq.~(\ref{eq:NPF}). We express the electric and magnetic
field in terms of the gauge field. After a Wick rotation to imaginary
time, the corresponding relations read $\:E_j=-\zeta A^j-i\pd_j A^0\:$
and $\:B_j=\varepsilon_{jkl}\pd_k A^l\:$ in Euclidean space.
Multiplying Eq.~(\ref{eq:EX2}) with $(\zeta \epsilon_\alpha)^{-1}$ and
decomposing
$\:\Lcal^\alpha_{j\mu}(\zeta;\xbf,\xbf')=\hat\nbf^\alpha_k(\xbf')
\Lcal^{k\alpha}_{j\mu}(\zeta;\xbf,\xbf')\:$ with respect to the
components $\hat\nbf^\alpha_k$ of the normal vector by using the
standard summation convention for $k$, one gets after some algebra the
explicit result
\begin{eqnarray}
\Lcal^{1\alpha}(\zeta;\xbf,\xbf') &=& 
\begin{pmatrix}
0 & -\frac{1}{\zeta\epsilon_\alpha}\big[
\bar{\pd}_3\pd_2-\bar{\pd}_2\pd_3\big] & 
\frac{1}{\zeta\epsilon_\alpha}\big[
\bar{\pd}_1\pd_3+\bar{\pd}_3\pd_1\big] & 
-\frac{1}{\zeta\epsilon_\alpha}\big[
\bar{\pd}_1\pd_2+
\bar{\pd}_2\pd_1\big]
\\[5pt]
i\pd_3 & 
-\frac{1}{\zeta\epsilon_\alpha}\bar{\pd}_1\pd_3 & 
\frac{1}{\zeta\epsilon_\alpha}\bar{\pd}_2\pd_3 & 
\zeta-\frac{1}{\zeta\epsilon_\alpha}\big[
\bar{\pd}_2\pd_2-\bar{\pd}_1\pd_1\big]
\\[5pt]
-i\pd_2 & 
\frac{1}{\zeta\epsilon_\alpha}\bar{\pd}_1\pd_2 & 
-\zeta+\frac{1}{\zeta\epsilon_\alpha}\big[
\bar{\pd}_3\pd_3-\bar{\pd}_1\pd_1\big] &
-\frac{1}{\zeta\epsilon_\alpha}\bar{\pd}_3\pd_2
\end{pmatrix}\,\Gcal_E^{\epsilon_\alpha}(\zeta;\xbf-\xbf')
\label{eq:DO1}\\[10pt]
\Lcal^{2\alpha}(\zeta;\xbf,\xbf') &=& 
\begin{pmatrix}
-i\pd_3 &
-\frac{1}{\zeta\epsilon_\alpha}\bar{\pd}_1\pd_3 & 
\frac{1}{\zeta\epsilon_\alpha}\bar{\pd}_2\pd_3 & 
-\zeta+\frac{1}{\zeta\epsilon_\alpha}\big[
\bar{\pd}_1\pd_1-\bar{\pd}_2\pd_2\big]
\\[5pt]
0 & -\frac{1}{\zeta\epsilon_\alpha}\big[
\bar{\pd}_2\pd_3+\bar{\pd}_3\pd_2\big] & 
-\frac{1}{\zeta\epsilon_\alpha}\big[
\bar{\pd}_1\pd_3-\bar{\pd}_3\pd_1\big] & 
\frac{1}{\zeta\epsilon_\alpha}\big[
\bar{\pd}_2\pd_1+\bar{\pd}_1\pd_2\big]
\\[5pt]
i\pd_1 & 
\zeta-\frac{1}{\zeta\epsilon_\alpha}\big[
\bar{\pd}_3\pd_3-\bar{\pd}_2\pd_2\big] & 
-\frac{1}{\zeta\epsilon_\alpha}\bar{\pd}_2\pd_1 & 
\frac{1}{\zeta\epsilon_\alpha}\bar{\pd}_3\pd_1 
\end{pmatrix}\,\Gcal_E^{\epsilon_\alpha}(\zeta;\xbf-\xbf')
\label{eq:DO2}\\[10pt]
\Lcal^{3\alpha}(\zeta;\xbf,\xbf') &=& 
\begin{pmatrix}
i\pd_2 & 
\frac{1}{\zeta\epsilon_\alpha}\bar{\pd}_1\pd_2 & 
\zeta-\frac{1}{\zeta\epsilon_\alpha}\big[
\bar{\pd}_1\pd_1-\bar{\pd}_3\pd_3\big] &
-\frac{1}{\zeta\epsilon_\alpha}\bar{\pd}_3\pd_2
\\[5pt]
-i\pd_1 & 
-\zeta+\frac{1}{\zeta\epsilon_\alpha}\big[
\bar{\pd}_2\pd_2-\bar{\pd}_3\pd_3\big] & 
-\frac{1}{\zeta\epsilon_\alpha}\bar{\pd}_2\pd_1 & 
\frac{1}{\zeta\epsilon_\alpha}\bar{\pd}_3\pd_1
\\[5pt]
0 & \frac{1}{\zeta\epsilon_\alpha}\big[
\bar{\pd}_3\pd_2+
\bar{\pd}_2\pd_3\big] & 
-\frac{1}{\zeta\epsilon_\alpha}\big[
\bar{\pd}_3\pd_1+
\bar{\pd}_1\pd_3\big] & 
-\frac{1}{\zeta\epsilon_\alpha}\big[
\bar{\pd}_2\pd_1-\bar{\pd}_1\pd_2\big]
\end{pmatrix}
\,\Gcal_E^{\epsilon_\alpha}(\zeta;\xbf-\xbf'),
\label{eq:DO3}
\end{eqnarray}
where $\:\Gcal_E^{\epsilon_\alpha}\:$ is the Euclidean version of the
Green function inside the medium which is defined by
Eq.~(\ref{eq:greenfunction}).  The partial differential operators
$\:\bar{\pd}_j\:$ are acting on the spatial argument of
$\:\Gcal_E^{\epsilon_\alpha}\:$, whereas the ``free'' operators
$\:\pd_j\:$ are acting on the gauge field to which
$\Lcal^\alpha_{j\mu}$ is applied. For non-deformed surfaces as
described by the conventional Lifshitz theory, i.e., $\hat \nbf
=(0,0,\pm 1)$, only the last matrix is relevant.

\subsection{General result for deformed surfaces}
\label{sec:general-result}

Now we are in the position to calculate the partition function defined
by Eq.~(\ref{eq:NPF}) and by the operators in
Eqs.~(\ref{eq:DO1}-\ref{eq:DO3}). Similar to the approach of Refs.
\cite{likar,golkar} we introduce auxiliary fields in order to treat
the delta-function constraints. However, here we will use complex
valued auxiliary fields since the arguments of the delta-functions are
complex in our problem. Moreover, the fields will be not defined on
the original surfaces $S_\alpha$ itself but on the flat auxiliary
surfaces $R_\alpha$ since these are the regions on which the
``external'' positions $\xbf$ of the boundary conditions are located,
cf.  Eq.~(\ref{eq:NPF}). Introducing on each of the two surfaces
$R_\alpha$ at $x_3=L_\alpha=(-1)^{\alpha-1}L$ with
lateral coordinates $\xbf_\|$ the three fields $\psi_{\alpha
  j}(\zeta,\xbf_\|)$, $(j=1,2,3)$, the delta-functions for fixed
$\alpha$ and $j$ can be written as
\begin{eqnarray}
&&
\prod_{\zeta_n} \prod_{\xbf \in R_\alpha} \delta\left[
\int_{\xbf' \in S_\alpha} \Lcal_{j\mu}^\alpha(\zeta_n;\xbf,\xbf') 
A^\mu(\zeta,\xbf')\right]\nonumber\\
&&\quad\quad =
\int \Dsc\left[\psi^\ast_{\alpha j} \psi_{\alpha j} \right]
\exp\left[ i \sum_n \int_{\xbf_\|} \int_{\xbf' \in S_\alpha}
\left\{
\psi^\ast_{\alpha j}(\zeta_n,\xbf_\|) 
\Lcal_{j \mu}^\alpha\left(\zeta_n; (\xbf_\|,L_\alpha),\xbf'\right)
A^\mu(\zeta_n,\xbf') + \text{c.c.}
\right\}\right].
\end{eqnarray}
Inserting this representation in the partition function of
Eq.~(\ref{eq:NPF}), the complex gauge field $A^\mu$ can be integrated
out, using the free action $S_E(A^\ast,A)$ of Eq.~(\ref{eq:action-e}).
The partition function can then be expressed in terms of an effective
quadratic action for the auxiliary fields,
\begin{equation}
\label{eq:Z-S-eff}
\Zcal^2(H)=\int \prod_{\alpha j}
\Dsc \left[\psi^\ast_{\alpha j} \psi_{\alpha j} \right]
e^{-S_\text{eff}[\psi^\ast_{\alpha j},\psi_{\alpha j}]}
\end{equation}
with
\begin{equation}
\label{eq:S-eff}
S_\text{eff}\left[\psi^\ast_{\alpha j},\psi_{\alpha j}\right]=
\sum_{n,n'} \int_{\xbf_\|}\int_{\xbf'_\|} 
\psi^\ast_{\alpha j}(\zeta_n,\xbf_{\|})
\Mcal^{\alpha\beta,jl}
(\zeta_n,\xbf_{\|};\zeta_{n'},\xbf_{\|}')\psi_{\beta l}
(\zeta_{n'},\xbf_{\|}'),
\end{equation}
where the usual summation convention applies to all indices. Since in
Feynman gauge the propagator of $A^\mu$ is diagonal in $\mu$, the
resulting matrix kernel can be simply written as
\begin{equation}\label{eq:kernel}
\Mcal^{\alpha\beta,jl}\big(\zeta,\xbf_{\|};\zeta',\xbf_{\|}'\big)\:=\:
2\pi\,\delta\big(\zeta-\zeta'\big) \int_{\ybf \in S_\alpha} 
\int_{\ybf' \in S_\beta}
\Lcal_{j\mu}^\alpha\left(\zeta;(\xbf_\|,L_\alpha),\ybf\right)
\,\Lcal_{\mu l}^{\dagger\beta}
\left(\zeta';(\xbf'_\|,L_\beta),\ybf'\right)\,\Gcal(\zeta;\ybf-\ybf'),
\end{equation}
where a summation over $\mu$ is implicit and $\Gcal(\zeta,\ybf)$ is
the free photon propagator with Fourier transform
$\Gcal(\zeta,\kbf)=(\zeta^2+\kbf^2)^{-1}$. To simplify this result,
and to prove the independence of the free energy on the
choice of $L$, it is useful to rewrite the operators of Eqs.
(\ref{eq:DO1}-\ref{eq:DO3}) as
$\:\Lcal^{k\alpha}(\zeta;\xbf,\ybf)\equiv \Lsc^{k\alpha}
(\zeta)\,\Gcal_E^{\epsilon_\alpha}(\zeta;\xbf-\ybf)\:$ in order to
make their proportionality to the material Green function explicit. It
is important to keep in mind that the differential operators $\Lsc^{k
  \alpha}$ act on the spatial arguments of $\Gcal_E^{\epsilon_\alpha}$
as well as on those of the free propagator $\Gcal$. Now the kernel in
Eq.~(\ref{eq:kernel}) acquires the form
\begin{eqnarray}\label{eq:kernel2}
\Mcal^{\alpha\beta,jl}\big(\zeta,\xbf_{\|};\zeta',\xbf_{\|}'\big)\:&=&\:
2\pi\,\delta\big(\zeta-\zeta'\big)
\int_{\ybf \in S_\alpha}\!\int_{\ybf' \in S_\beta}\!
\hat{\nbf}^\alpha_k\,
\hat{\nbf}'^\beta_s\:
\big[\Lsc^{k\alpha}\!\cdot\Lsc'^{\dagger s\beta}\big]_{jl}\nonumber\\
\quad &&\:\times\:
\Gcal_E^{\epsilon_\alpha}(\zeta;\xbf-\ybf)_{|x_3=L_\alpha}\,
\Gcal_E^{\epsilon_\beta}(\zeta';\xbf'-\ybf')_{|x'_3=L_\beta}\,
\Gcal(\zeta;\ybf-\ybf'),
\end{eqnarray}
where $\:\Lsc'^{k\alpha}$ acts on the primed coordinates and a
summation over $k$ and $s$ is implicit. In momentum space, using
$\Gcal_E^\epsilon(\zeta,\kbf)=(\zeta^2 \epsilon(i\zeta)+\kbf^2)^{-1}$,
the partially Fourier transformed material Green function can be written as
\begin{equation}
  \label{eq:part-Green-medium}
  \Gcal^{\epsilon_\alpha}_E(\zeta;\kbf_\|,z)=\frac{e^{-p_\alpha(\zeta,
\kbf_\|)|z|}}
{2p_\alpha(\zeta,\kbf_\|)}
\end{equation}
with
$p_\alpha(\zeta,\kbf_\|)=\sqrt{\epsilon_\alpha(i\zeta)\zeta^2+\kbf_\|^2}$.
With this representation the kernel becomes
\begin{equation}
\begin{split}
\Mcal^{\alpha\beta,jl}\big(\zeta,\kbf_\|;\zeta',\kbf_\|'\big)
\:=\:2\pi\,\delta(\zeta-\zeta')
&\int_{\ybf\in S_\alpha}\int_{\ybf'\in S_\beta}
\,e^{-i\kbf_\|\ybf_\|+i\kbf_\|'\ybf_\|'}\,
\frac{e^{-p_\alpha(\zeta,\kbf_\|)\abs{L_\alpha-y_3}}}
{2p_\alpha(\zeta,\kbf_\|)}
\frac{e^{-p_\beta(\zeta',\kbf'_\|)\abs{L_\beta-y'_3}}}
{2p_\beta(\zeta',\kbf'_\|)}\\
&\times 
\hat{\nbf}^\alpha_k
\,\hat{\nbf}'^\beta_s\,\big[\hat{\Lsc}^{k\alpha}(\zeta,\kbf_\|)
\!\cdot\hat{\Lsc}'^
{\dagger s\beta}(\zeta',\kbf'_\|)\big]_{jl}\,
\Gcal(\zeta;\ybf-\ybf'),
\end{split}
\end{equation}
where the differential operators $\hat{\Lsc}^{k\alpha}(\zeta,\kbf_\|)$
are obtained from the $\Lsc^{k\alpha}$ of
Eqs.~(\ref{eq:DO1}-\ref{eq:DO3}) by the replacements $\bar\nabla_\|
\equiv (\bar\pd_1,\bar\pd_2) \ra i\kbf_\|$, $\bar{\pd}_3\ra(-1)^\alpha
p_\alpha$.  Thus the operators $\hat{\Lsc}^{k\alpha}(\zeta,\kbf_\|)$
are acting via the remaining derivatives $\pd_j$ only on the spatial
coordinates of the vacuum Green function $\Gcal(\zeta;\ybf)$.  At this
stage it will become obvious that the free energy or force is
independent of the positions $x_3= \pm L$ of the auxiliary surfaces
$R_\alpha$.  Due to the construction of the surfaces $R_\alpha$, we
have $\abs{L_\alpha-y_3}=(-1)^{\alpha-1}(L_\alpha-y_3)$.  A
consequence of this is the important observation that the kernel can
be factorized into
\begin{equation}
\label{M-factor}
\Mcal^{\alpha\beta,jl}\big(\zeta,\kbf_\|;\zeta',\kbf_\|'\big)\:=\:
\eta_\alpha(\zeta,\kbf_\|)\,
\til{\Mcal}^{\alpha\beta,jl}\big(\zeta,\kbf_\|;\zeta',\kbf_\|'\big)\,
\eta_\beta(\zeta',\kbf_\|')
\end{equation}
with the functions $\eta_\alpha(\zeta,\kbf_\|)=\exp(-
p_\alpha(\zeta,\kbf_\|)L )/2p_\alpha(\zeta,\kbf_\|)$ and the simplified
$L$-independent kernel
\begin{equation}
\label{eq:M-simple}
\begin{split}
\til{\Mcal}^{\alpha\beta,jl}
\big(\zeta,\kbf_\|;\zeta',\kbf_\|'\big)\:=\:2\pi\,\delta(\zeta-\zeta')
&\int_{\ybf\in S_\alpha}\int_{\ybf'\in S_\beta}
\,e^{-i\kbf_\|\ybf_\|+i\kbf_\|'\ybf_\|'}\,
e^{-[(-1)^\alpha p_\alpha(\zeta,\kbf_\|) y_3
+(-1)^\beta p_\beta(\zeta',\kbf'_\|) y'_3]}\\
&\times\,
\hat{\nbf}^\alpha_k
\,\hat{\nbf}'^\beta_s\,\big[\hat{\Lsc}^{k\alpha}(\zeta,\kbf_\|)
\!\cdot\hat{\Lsc}'^
{\dagger s\beta}(\zeta',\kbf'_\|)\big]_{jl}\,
\Gcal(\zeta;\ybf-\ybf').
\end{split}
\end{equation}
From Eq.~(\ref{eq:Z-S-eff}) follows that the partition function
$\Zcal(H)=\det^{-1/2} \Mcal$ with the determinant taken with respect
to both the continuous ($\zeta$, $\kbf_\|$) and the discrete
($\alpha$, $j$) arguments. Due to the structure of
Eq.~(\ref{M-factor}) one has $\det \Mcal \propto \det \til\Mcal$.
Since the functions $\eta_\alpha(\zeta,\kbf_\|)$ are independent of
the mean surface distance $H$, the proportionality constant of the two
determinants is independent of $H$, too. Therefore, this
constant, and the $L$-dependence, will drop out of the
free energy of Eq.~(\ref{def-energy}) which can now be written as
\begin{equation}
  \label{eq:free-energy-gen}
  \Fcal(H)=\frac{1}{2A\beta} \ln \det \left( \til\Mcal 
\til\Mcal_\infty^{-1} \right),
\end{equation}
where $\til\Mcal_\infty$ denotes $\til\Mcal$ in the limit of
asymptotically large $H$. The force per unit area between the two
surfaces can be directly obtained from the kernel $\til\Mcal$ by
\begin{equation}
  \label{eq:force-gen}
  F=-\frac{1}{2A\beta} \text{Tr} \left(\til\Mcal^{-1} \pd_H \til\Mcal \right)
\end{equation}
without the need to subtract the asymptotic expansion for large $H$.
Here the trace has to be taken with respect to the Matsubara
frequencies $\zeta_n$, the lateral momenta $\kbf_\|$, and the discrete
indices $\alpha$, $j$.  Eqs.~(\ref{eq:free-energy-gen}) and
(\ref{eq:force-gen}) together with Eq.~(\ref{eq:M-simple}) represents
the main result of our general approach. We will apply this formula
below to specific model situations. Before proceeding along these
lines, it might be interesting to discuss the above result. During the
derivation of the result we worked within the Feynman gauge. This,
however, poses no problem since the restricted partition function
$\Zcal(H)$ can be considered as the expectation value of the boundary
condition enforcing delta-functions with respect to the free action of
the gauge field. The arguments of the delta-functions are composed of
the electromagnetic field components, and are thus manifestly gauge
invariant. Therefore, the kernel $\til\Mcal$ must be gauge invariant.

Let us first discuss the most simple situation where the kernel
$\til\Mcal$ is diagonal in momentum space so that the force can be
calculated exactly. This will be the case when the geometry has
translational symmetry in the lateral directions, i.e., for flat
surfaces. Then the integrals in Eq.~(\ref{eq:M-simple}) can be easily
computed, and the resulting kernel provides a compact account of
Lifshitz theory as we will show in the next Section. Even if the
surfaces are deformed the kernel can be obtained explicitly if one
considers the limit of ideal metals, i.e., a diverging dielectric
function $\epsilon(i\zeta)$. In this particular limit both $p_\alpha$
and the operators $\hat \Lsc^{k\alpha}$ become independent of the
lateral momentum $\kbf_\|$. Therefore, after parameterizing the
surfaces so that $y_3$, $y'_3$ are replaced by functions of the
lateral coordinates $\ybf_\|$, $\ybf'_\|$, respectively, the integrals
in Eq.~(\ref{eq:M-simple}) correspond to Fourier transformations with
respect to the lateral coordinates, and the kernel assumes a simple
form in position space as we will demonstrate explicitly below.
However, any kind of deviation from flat surfaces (even for ideal
metals) renders $\til\Mcal$ non-diagonal and makes the evaluation of
Eq.~(\ref{eq:force-gen}) a hard problem. There are basically two
approaches to tackle this problem. First, one can consider the
amplitude of the surface deformations as small compared to both the
mean surface distance and other characteristic lateral length scales
as, e.g., the roughness correlation length. Then one can apply
perturbation theory to obtain the force in powers of the deformation
profiles $h_\alpha(\xbf_\|)$. This program has been carried out in
detail for ideal metals in Refs. \cite{Emig+01,Emig+03}. Second, one
can try to compute the force exactly by a numerical algorithm. For
periodically deformed (corrugated) surfaces of ideal metals it has
been demonstrated recently that the corresponding kernel can be
transformed to a form which is particularly suited for an efficient
numerical evaluation of the force \cite{emig}. We expect that these
techniques can be applied to the general case of deformed surfaces of
dielectric media using our approach of expressing the force in terms
of a kernel [Eq.~(\ref{eq:M-simple})] which contains all information
about material and geometrical properties of the surfaces. As for
ideal metals, the kernel is proportional to the vacuum Green function
which, however, is now dressed by the operators $\hat \Lsc^{k \alpha}$
which contain the reflection and refraction properties of the
material.

\section{Flat Surfaces}
\label{sec:lt}

As a simple application of our approach, we consider in this section
the case of flat surfaces of general dielectric media. In this
particular limit the force between the surfaces is well-known from
more conventional approaches.  The corresponding result is known as
the so-called Lifshitz theory for molecular forces \cite{Lifshitz}. In
the following we will show that our path-integral approach provides a
compact derivation of the Lifshitz result without the need to solve
Maxwell's equations with a random source explicitly and to calculate
the expectation value of the stress tensor. In the flat surface limit,
the surfaces $S_\alpha$ are parameterized by $(\ybf_\|,H_\alpha)$ with
$H_\alpha=0$, $H$ for $\alpha=1$, $2$, respectively. Due to the
translational symmetry of the problem, it is convenient to work in
momentum space. Using the representation
\begin{equation}
  \label{eq:G-0-FT}
  \Gcal(\zeta,\ybf)=\int_{\qbf_\|} e^{i \qbf_\| \ybf_\|} 
\frac{e^{-p(\zeta,\qbf_\|)|y_3|}}{2p(\zeta,\qbf_\|)}
\end{equation}
of the vacuum Green function in Eq.~(\ref{eq:M-simple}) with
$p(\zeta,\qbf_\|)=\sqrt{\zeta^2+\qbf_\|^2}$ yields
\begin{equation}\label{eq:flat_limit}
\begin{split}
\tilde\Mcal^{\alpha\beta,jl}\big(\zeta,\kbf_\|;\zeta',\kbf_\|'\big)\:=\:
2\pi\,\delta(\zeta-\zeta')\int_{\qbf_\|}&\int_{\ybf_\|}\!
\int_{\ybf'_\|}\!
e^{-i\kbf_\|\ybf_\|+i\kbf_\|'\ybf_\|'}\,
e^{-[(-1)^\alpha p_\alpha(\zeta,\kbf_\|) H_\alpha
+(-1)^\beta p_\beta(\zeta',\kbf'_\|)H_\beta]}\,(-1)^{\alpha+\beta}
\\[5pt]
&\times\,
\big[\hat{\Lsc}^{3\alpha}(\zeta,\kbf_\|)\!
\cdot\hat{\Lsc}'^{\dagger3\beta}(\zeta',\kbf'_\|)\big]_{jl}
e^{i\qbf_\|(\ybf_\|-\ybf_\|')}
\frac{e^{-p(\zeta,\qbf_\|)\abs{y_3-y_3'}}}{2p(\zeta,\qbf_\|)},
\end{split}
\end{equation}
where we made use of the surface normal vectors $\hat\nbf^\alpha
=(0,0,(-1)^{\alpha-1})$ for flat surfaces. The differential operators
$\hat{\Lsc}^{3\alpha}$ and $\hat{\Lsc}'^{\dagger3\beta}$ can now
be expressed in momentum space with the replacements $\nabla_\|\ra
i\kbf_\|$ and $\nabla_\|'\ra i\kbf_\|'$ yielding
\begin{equation}
\tilde\Mcal^{\alpha\beta,jl}\big(\zeta,\kbf_\|;\zeta',\kbf_\|'\big)\:=\:
(2\pi)^3\delta(\zeta-\zeta')\,\delta^{(2)}\big(\kbf_\|-\kbf_\|'\big)\,
\eta^{\alpha\beta}\,
\big[\hat{\Lsc}^{3\alpha}(\zeta,\kbf_\|)\!\cdot\hat{\Lsc}'^{\dagger3\beta}
(\zeta',\kbf'_\|)\big]_{jl}
\frac{e^{-p(\zeta,\kbf_\|)\abs{y_3-y_3'}}}{2p(\zeta,\kbf_\|)}\Big\vert_{
\begin{smallmatrix}
y_3=H_\alpha\\
y_3'=H_\beta
\end{smallmatrix}}
\end{equation}
where we separated the factor $\eta^{\alpha\beta}=(-1)^{\alpha+\beta}
e^{-[(-1)^\alpha p_\alpha(\zeta,\kbf_\|) H_\alpha +(-1)^\beta
  p_\beta(\zeta',\kbf'_\|)H_\beta]}$ which we will discuss below. The
differential operator acquires now the form
\begin{equation}
\hat{\Lsc}^{3\alpha}\big(\zeta,\kbf_\|\big)\:=\: 
\begin{pmatrix}
-k_2 & 
-\frac{k_1k_2}{\zeta\epsilon_\alpha} & 
\zeta+\frac{1}{\zeta\epsilon_\alpha}\big[
k_1^2+(-1)^\alpha p_\alpha\pd_3\big] &
-i\frac{(-1)^\alpha p_\alpha}{\zeta\epsilon_\alpha}k_2
\\[5pt]
k_1 & 
-\zeta-\frac{1}{\zeta\epsilon_\alpha}\big[
k_2^2+(-1)^\alpha p_\alpha\pd_3\big] & 
\frac{k_1k_2}{\zeta\epsilon_\alpha} & 
i\frac{(-1)^\alpha p_\alpha}{\zeta\epsilon_\alpha}k_1
\\[5pt]
0 & \frac{ik_2}{\zeta\epsilon_\alpha}\big[
(-1)^\alpha p_\alpha+\pd_3\big] & 
-\frac{ik_1}{\zeta\epsilon_\alpha}\big[
(-1)^\alpha p_\alpha+\pd_3\big] & 0
\end{pmatrix},
\end{equation}
and the primed adjoint operator acts via $\partial_3$ on $y'_3$.
Before we calculate from this expression the free energy and force
between the surfaces, it is instructive to examine the structure of
the above matrix. It is easily checked that the third row of the
matrix--operator $\hat{\Lsc}^{3\alpha}$ can be expressed in terms of
the other two rows via $(-1)^\alpha
ip_\alpha\hat{\Lsc}^{3\alpha}_{3\mu}
=k_1\hat{\Lsc}^{3\alpha}_{1\mu}+k_2\hat{\Lsc}^{3\alpha}_{2\mu}$.  The
physical reason for this can be easily understood. The relation
between the rows reflects the fact that there exist only two
independent boundary conditions for each surface. Since the surfaces
are flat here, any field configurations can be considered as a
superposition of transversal electric (TE) and magnetic (TM) modes.
Each mode type is characterized by a scalar field which has to satisfy
only one boundary condition at each surface. This will be demonstrated
more explicitly in the Appendix where the problem is formulated from
the outset in terms of two scalar fields representing TE and TM modes.
However, for deformed surfaces this reduction to separated mode types
is generally no longer expected to hold since the modes will mix under
the scattering at deformations.

For flat surfaces we are thus led to introduce the reduced
matrix--operator
\begin{equation}\label{eq:EDO}
\Ombf^\alpha(\zeta,\kbf_\|;\pd_z)\:=\:
\begin{pmatrix}
-k_2 & -\frac{k_1k_2}{\zeta\epsilon_\alpha} & 
\zeta+\frac{1}{\zeta\epsilon_\alpha}
[k_1^2+(-1)^\alpha p_\alpha\,\pd_z] & 
-i\frac{(-1)^\alpha p_\alpha k_2}{\zeta\epsilon_\alpha}
\\[5pt]
k_1 & -\zeta-\frac{1}{\zeta\epsilon_\alpha}
[k_2^2+(-1)^\alpha p_\alpha\,\pd_z] & 
\frac{k_1k_2}{\zeta\epsilon_\alpha} & 
i\frac{(-1)^\alpha p_\alpha k_1}{\zeta\epsilon_\alpha}
\end{pmatrix}
\end{equation}
which consists of two linear independent rows only. Defining
$\Ombf^\alpha_\pm(\zeta,\kbf_\|)\equiv\Ombf^\alpha(\zeta,\kbf_\|;\pm\pd_z)$
the kernel can be written as
\begin{equation}\label{eq:kernel4}
\tilde\Mcal^{\alpha\beta,jl}\big(\zeta,\kbf_\|;\zeta',\kbf_\|'\big)\:=\:
(2\pi)^3 \delta(\zeta-\zeta')
\delta^{(2)}\big(\kbf_\|-\kbf_\|'\big)\,\eta^{\alpha\beta}\,
\Big[\big(\Ombf^\alpha_+(\zeta,\kbf_\|)\!\cdot\Ombf^{\dagger\beta}_-
(\zeta',\kbf'_\|)\big)_{jl}
\frac{e^{-p(\zeta,\kbf_\|)\abs{z}}}{2p(\zeta,\kbf_\|)}
\Big]_{z=H_\alpha-H_\beta}.
\end{equation}  
The entries of this diagonal matrix consist of the $4\times 4$
matrices which are given by the expression in the square brackets.
Inserting now Eq.~(\ref{eq:EDO}) into Eq.~(\ref{eq:kernel4}) we obtain
for the expression in the square brackets the $4 \times 4$ matrix
\begin{equation}\label{eq:kernel5}
\tilde\Mcal\big(\zeta,\kbf_{\|}\big)\:=\:
\twomatrix{A_1}
{\frac{e^{-pH}}{\epsilon_1 \epsilon_2 \zeta^2} B}
{\frac{e^{-pH}}{\epsilon_1 \epsilon_2 \zeta^2} B}
{A_2}
\end{equation}
in terms of the symmetric $2 \times 2$ matrices 
\begin{equation}
A_\alpha \: = \: \twomatrix
{\frac{\epsilon_\alpha-1}{\epsilon_\alpha} \zeta^2+
\frac{\epsilon_\alpha-1}{\epsilon_\alpha^2} k_1^2+
\frac{\epsilon_\alpha^2-1}{\epsilon_\alpha^2} k_2^2}
{-\frac{\epsilon_\alpha-1}{\epsilon_\alpha}k_1 k_2}
{-\frac{\epsilon_\alpha-1}{\epsilon_\alpha}k_1 k_2}
{\frac{\epsilon_\alpha-1}{\epsilon_\alpha} \zeta^2+
\frac{\epsilon_\alpha^2-1}{\epsilon_\alpha^2} k_1^2+
\frac{\epsilon_\alpha-1}{\epsilon_\alpha^2} k_2^2},
\end{equation}
\begin{equation}
B\: = \: 
\:\twomatrix
{(p_1^2 - k_2^2-pp_1)(p_2^2 -k_2^2-pp_2)+
k_2^2(\epsilon_1\epsilon_2\zeta^2+k_1^2-p_1p_2)}
{k_1 k_2\left( (\epsilon_1+\epsilon_2-\epsilon_1\epsilon_2)\zeta^2 
+\kbf_\|^2 -p(p_1+p_2)+p_1p_2\right)}
{k_1 k_2\left( (\epsilon_1+\epsilon_2-\epsilon_1\epsilon_2)\zeta^2 
+\kbf_\|^2 -p(p_1+p_2)+p_1p_2\right)}
{(p_1^2 - k_1^2 -pp_1)(p_2^2 -k_1^2-pp_2)+
k_1^2(\epsilon_1\epsilon_2\zeta^2+k_2^2-p_1p_2)}
\end{equation}
using $p_\alpha=\sqrt{\epsilon_\alpha\zeta^2+\kbf_\|^2}$,
$p=\sqrt{\zeta^2+\kbf_\|^2}$. The Casimir free energy per unit area
can now be obtained from Eq.~(\ref{eq:free-energy-gen}). In the limit
$H\to\infty$ the off-diagonal elements of $\tilde\Mcal$ in
Eq.~(\ref{eq:kernel5}) vanish, so that we have to compute the
determinant of the matrix
\begin{equation}
\label{eq:red-lifshitz-kernel}
\tilde\Mcal(\zeta,\kbf_\|)\tilde\Mcal^{-1}_\infty(\zeta,\kbf_\|)=
\twomatrix
{I}
{\frac{e^{-pH}}{\epsilon_1 \epsilon_2 \zeta^2} B A_2^{-1}}
{\frac{e^{-pH}}{\epsilon_1 \epsilon_2 \zeta^2} B A_1^{-1}}
{I},
\end{equation}
where $I$ is the $2\times 2$ identity matrix. In the above matrix we
have neglected the factor $\eta_{\alpha\beta}$ appearing in
Eq.~(\ref{eq:kernel4}). This factor will have no effect on the free
energy as will show at the end of this section. The determinant of the
matrix of Eq.~(\ref{eq:red-lifshitz-kernel}) can be calculated using
the relation
\begin{equation}
  \label{eq:gen-det}
  \det(Y)=1-\text{Tr}(X_1 X_2) +\det(X_1 X_2) 
\end{equation}
for a general $4 \times 4$ matrix of the form
\begin{equation}
  \label{eq:gen-matrix}
  Y=\twomatrix{I}{X_1}{X_2}{I}.
\end{equation}
Thus the free energy can be obtained by calculating the determinant of
just a $2 \times 2$ matrix. Using Eq. (\ref{eq:free-energy-gen}), the
logarithm of the product of all the determinants for different
$\zeta_n$ and $\kbf_\|$ becomes a corresponding sum and integral,
respectively,
\begin{equation}
\Fcal(H)=\frac{1}{2\beta}\sum_{n=-\infty}^\infty \int_{\kbf_\|}
\ln \abs{\tilde\Mcal(\zeta_n,\kbf_\|)\tilde\Mcal^{-1}_\infty(\zeta_n,\kbf_\|)},
\end{equation}
where $|\ldots |$ denotes the determinant of the $4 \times 4$ matrix
at fixed $\zeta$ and $\kbf_\|$. Calculating explicitly the determinant
with the aid of Eq.~(\ref{eq:gen-det}), we obtain the final result for
the Casimir or interaction free energy per unit area of the surfaces,
\begin{equation}\label{eq:energie}
\Fcal(H)\:=\:\frac{1}{\beta}\sideset{}{'}\sum_{n=0}^{\infty}
\int_{0}^{\infty}\!\frac{k\,dk}{2\pi}\:\ln\Bigg(
\bigg[1-e^{-2p_nH}\frac{p_{n1}-p_n}{p_{n1}+p_n}\,
\frac{p_{n2}-p_n}{p_{n2}+p_n}\bigg]
\bigg[1-e^{-2p_nH}\frac{p_{n1}-\epsilon_1p_n}{p_{n1}+\epsilon_1p_n}\,
\frac{p_{n2}-\epsilon_2p_n}{p_{n2}+\epsilon_2p_n}\bigg]\Bigg).
\end{equation}
with $k=|\kbf_\| |$. The corresponding force per unit area is given by
\begin{equation}\label{eq:kraft}
F(H)\:=\:
-\frac{1}{\beta}\sideset{}{'}\sum_{n=0}^{\infty}\!
\int_{0}^{\infty}\!\frac{kdk}{\pi}\,p_n\:\Bigg(
\bigg[\frac{p_{n1}+p_n}{p_{n1}-p_n}\,
\frac{p_{n2}+p_n}{p_{n2}-p_n}\,e^{2p_nH}-1\bigg]^{-1}\!\!+\:
\bigg[\frac{p_{n1}+\epsilon_1p_n}{p_{n1}-\epsilon_1p_n}\,
\frac{p_{n2}+\epsilon_2p_n}{p_{n2}-\epsilon_2p_n}\,e^{2p_nH}-1
\bigg]^{-1}\Bigg),
\end{equation}
where we defined
$p_{n\alpha}=\sqrt{\epsilon_\alpha\zeta_n^2+k^2}$ and
$p_n=\sqrt{\zeta_n^2+k^2}$.  The primed sum indicates
that the term for $\:n=0\:$ is to be multiplied by $1/2$. It is
important to note that the dielectric function in the above
expressions is evaluated along the imaginary axis only, since
$\epsilon_\alpha\equiv \epsilon_\alpha(i\zeta)$ due to the initial
Wick rotation to the Euclidean field theory. Since
$\epsilon_\alpha(i\zeta)$ is completely determined by the imaginary
part of the dielectric function for real frequencies $\omega$, the
force depends only the dissipative properties of the media, as
expected from the fluctuations-dissipation theorem. Our result of
Eqs.~(\ref{eq:energie}), (\ref{eq:kraft}) is in perfect agreement with
the original result by Lifshitz \cite{Lifshitz,milonni}.

Next we consider now the zero temperature limit. This
limit is obtained by the replacements $\zeta_n\ra\zeta$ and
$1/\beta\sum_{n\ge0}'\ra\int_0^\infty d\zeta/2\pi$ in Eqs.
(\ref{eq:energie}) and (\ref{eq:kraft}). Following Lifshitz, we
change the integration variable to $q=\sqrt{1+k^2/\zeta^2}$ and define
$s_\alpha\equiv\sqrt{q^2-1+\epsilon_\alpha(i\zeta)}$, yielding
\begin{equation}\label{eq:lifshitzenergy2}
\Fcal(H)\:=\:
\int_{0}^{\infty}\!\frac{\zeta^2d\zeta}{2\pi}
\int_{1}^{\infty}\!\frac{q\,dq}{2\pi}\:\ln\Bigg(
\bigg[1-e^{-2\zeta qH}\frac{s_1-q}{s_1+q}\,
\frac{s_2-q}{s_2+q}\bigg]
\bigg[1-e^{-2\zeta qH}\frac{s_1-\epsilon_1q}{s_1+\epsilon_1q}\,
\frac{s_2-\epsilon_2q}{s_2+\epsilon_2q}\bigg]\Bigg)
\end{equation}
for the free energy, and
\begin{equation}\label{eq:lifshitzforce}
F(H)\:=\:
-\frac{1}{2\pi^2}\int_{0}^{\infty}\!\zeta^3d\zeta
\int_{1}^{\infty}\!q^2dq\:\Bigg(
\bigg[\,\frac{s_1+q}{s_1-q}\,\frac{s_2+q}{s_2-q}
\:e^{2\zeta qH}-1\,\bigg]^{-1}
\:+\:\bigg[\,\frac{s_1+\epsilon_1q}{s_1-\epsilon_1q}\,
\frac{s_2+\epsilon_2q}{s_2-\epsilon_2q}
\:e^{2\zeta qH}-1\,\bigg]^{-1}\Bigg)
\end{equation}
for the force. The above result agrees again with the original
Lifshitz theory, see Eq. (2.9) in \cite{Lifshitz}.

Finally, we come back to the omitted factor $\eta_{\alpha\beta}$. The
effect of taking into account this factor is that in the matrix-kernel
of Eq.~(\ref{eq:kernel5}) the matrix $A_2$ is multiplied by
$e^{-2p_2H}$, $B$ is multiplied by $-e^{-p_2H}$ while $A_1$ remains
unchanged. For the matrix in Eq.~(\ref{eq:red-lifshitz-kernel}) this
means that the off-diagonal matrix $\sim A_2^{-1}$ gets multiplied by
the factor $-e^{p_2H}$ while the matrix $\sim A_1^{-1}$ gets
multiplied by the inverse factor $-e^{-p_2H}$. Due to
Eq.~(\ref{eq:gen-det}) the determinant depends only on the product of
the two off-diagonal matrices so that the factors coming from the
$\eta_{\alpha\beta}$ drop out in the determinant of $\tilde\Mcal
\tilde\Mcal^{-1}_\infty$.

\section{Deformed surfaces of ideal metals}
\label{sec:ideal-metal}

In the previous section we saw that our general approach reproduces
the Lifshitz theory for {\it flat} surfaces of dielectric media. In
this section we will apply our theory to deformed surfaces. As an
example we consider ideal metals with infinite dielectric functions
$\epsilon_\alpha$. This is a reasonable approximation for surface
separations which are large compared to the plasma wavelength of the
material. However, our general result for the kernel of
Eq.~(\ref{eq:M-simple}) contains all information which is necessary to
treat deformed surfaces of non-ideal metals or general dielectric
media as well. In the latter case the kernel $\Mcal$ assumes in
general no particular simple form and has to be computed numerically
in order to obtain the force. For ideal metals the kernel can be
calculated explicitly and the result provides another interesting
limit which has not been studied previously. In previous works only
special deformations of ideal metals have been studied by a path
integral approach. If the surface deformations are translational
invariant in one direction as for, e.g. uni-axial corrugations, the
electromagnetic field can be separated into TE and TM modes. This
property has been used in \cite{Emig+01,Emig+03,emig} to describe the
surface interaction by a scalar field theory. In contrast, here we
will allow for general deformations so that no separation into TE and
TM modes is possible anymore.

Our starting point is the general result for the kernel of
Eq.~(\ref{eq:M-simple}). After taking the limit $\epsilon_\alpha \to
\infty$ both $p_\alpha(\zeta,\kbf_\|)$ and the operators $\hat
\Lsc_{k\alpha}$ become independent of the lateral momentum $\kbf_\|$.
The kernel can then be written as
\begin{equation}\label{eq:high-eps-limit}
\begin{split}
\til{\Mcal}^{\alpha\beta,jl}\big(\zeta,\kbf_\|;\zeta',\kbf_\|'\big)
\:=\:
2\pi\,\delta(\zeta-\zeta')&\int_{\ybf\in S_\alpha}\!
\int_{\ybf'\in S_\beta}\!
e^{-i\kbf_\|\ybf_\|-i\kbf_\|'\ybf_\|'}\,
e^{-\abs{\zeta}\,[(-1)^\alpha\sqrt{\epsilon_\alpha}\,y_3
+(-1)^\beta\sqrt{\epsilon_\beta}\,y_3']}
\\[5pt]
&\times\,\hat{\nbf}^\alpha_k\,\hat{\nbf}'^\beta_s\,
\big[\hat{\Lsc}^{k\alpha}(\zeta)
\!\cdot\hat{\Lsc}'^{\dagger s\beta}(\zeta')\big]_{jl}
\Gcal\big(\zeta;\ybf-\ybf'\big)
\end{split}
\end{equation}
with the differential operators
\begin{equation}\label{eq:big_epsilon_DO}
\hat{\Lsc}^{1\alpha}\:=\:
\begin{pmatrix}
0 & 0 & 0 & 0\\[5pt]
i\pd_3 & 0 & 0 & \zeta\\[5pt]
-i\pd_2 & 0 & -\zeta & 0
\end{pmatrix}
\qquad
\hat{\Lsc}^{2\alpha}\:=\: 
\begin{pmatrix}
-i\pd_3 & 0 & 0 & -\zeta\\[5pt]
0 & 0 & 0 & 0\\[5pt]
i\pd_1 & \zeta & 0 & 0 
\end{pmatrix}
\qquad
\hat{\Lsc}^{3\alpha}\:=\: 
\begin{pmatrix}
i\pd_2 & 0 & \zeta & 0\\[5pt]
-i\pd_1 & -\zeta & 0 & 0\\[5pt] 
0 & 0 & 0 & 0
\end{pmatrix}.
\end{equation}
Due to the simple exponential dependence of the integrand of
Eq.~(\ref{eq:high-eps-limit}) on $\kbf_\|$ it is more convenient to
transform the kernel to position space.  When we insert the height
profile of the surfaces with $y_3=H_\alpha+h_\alpha(\ybf_\|)$,
$y'_3=H_\beta+h_\beta(\ybf'_\|)$, $H_\alpha=0$, $H$ for $\alpha=1$,
$2$, the position space form of the kernel can be read off from
Eq.~(\ref{eq:high-eps-limit}),
\begin{equation}\label{eq:big_epsilon_kernel}
\begin{split}
\til{\Mcal}^{\alpha\beta,jl}\big(\zeta,\ybf_{\|};\zeta',\ybf_{\|}'\big)
\:=\:2\pi\,\delta(\zeta-\zeta')\,&
e^{-\abs{\zeta}\,[(-1)^\alpha\sqrt{\epsilon_\alpha}\,H_\alpha
+(-1)^\beta\sqrt{\epsilon_\beta}\,H_\beta]}\:
e^{-\abs{\zeta}\,[(-1)^\alpha\sqrt{\epsilon_\alpha}\,h_\alpha(\ybf_{\|})
+(-1)^\beta\sqrt{\epsilon_\beta}\,h_\beta(\ybf_{\|}')]}\\
&\times\,
\hat{\nbf}^\alpha_k\,\hat{\nbf}'^\beta_s\,
\big[\hat\Lsc^{k\alpha}(\zeta)\!\cdot\hat\Lsc'^{\dagger s\beta}
(\zeta')\big]_{jl}
\Gcal\big(\zeta;\ybf-\ybf'\big)\bigg\vert_
{\begin{smallmatrix}
y_3=H_\alpha+h_\alpha(\ybf_{\|})\\
y_3'=H_\beta+h_\beta(\ybf_{\|}')
\end{smallmatrix}}.
\end{split}
\end{equation}
Before proceeding, it is useful to discuss the two exponential factors
depending on $\sqrt{\epsilon_\alpha}$. Let us start with the second
one which depends on the height profiles $h_\alpha(\ybf_\|)$ but is
independent of the mean surface distance $H$. Defining
$\eta_\alpha(\zeta,\ybf_\|)=e^{-\abs{\zeta}
  (-1)^\alpha\sqrt{\epsilon_\alpha}\,h_\alpha(\ybf_{\|})}$ the matrix
has the same structure as in Eq.~(\ref{M-factor}) but with $\kbf_\|$
replaced by $\ybf_\|$. Due to the arguments given below
Eq.~(\ref{M-factor}) the factors $\eta_\alpha(\zeta,\ybf_\|)$ drop out
of the free energy and can thus be neglected in the following. For the
first exponential factor in Eq.~(\ref{eq:big_epsilon_kernel}) this
argument does not apply since the factor depends on $H$. However, we
can make use of the fact that the factor does not depend on the
lateral coordinates $\ybf_\|$. The effect of this exponential factor
is that every $2\times 2$ sub-matrix of $\tilde \Mcal$ resulting from
keeping $(j,l)$, $(\zeta,\ybf_\|)$ and $(\zeta',\ybf'_\|)$ fixed is
multiplied by the same only $\zeta$ dependent factors.  The two
diagonal elements are multiplied by $1$ and
$e^{-2|\zeta|\sqrt{\epsilon_2}H}$, respectively, while the
off-diagonal elements are multiplied by
$e^{-|\zeta|\sqrt{\epsilon_2}H}$. It is easy to check that this leads
to the simple factor $e^{-N|\zeta|\sqrt{\epsilon_2}H}$ for the
determinant of $\tilde \Mcal$ if $N$ denotes the dimension of the
matrix $\tilde\Mcal$. However, this factor will drop out when taking
the determinant of the ratio $\tilde \Mcal \tilde \Mcal_\infty^{-1}$.
Therefore, the first exponential factor in Eq.
(\ref{eq:big_epsilon_kernel}) can be omitted as well.

Now the kernel assumes a simple form. Expressing the surface normal
vectors in terms of the height profile by
\begin{equation}
\label{eq:normal-vector}
\hat{\nbf}_\alpha\:=\:
\frac{(-1)^\alpha}{\sqrt{g_\alpha}}
\begin{pmatrix}
h_{\alpha,1}\\[5pt]
h_{\alpha,2}\\[5pt]
-1
\end{pmatrix}
\end{equation}
with $h_{\alpha,j}=\pd_jh_\alpha$ and $g_\alpha\:=\:
1+\big(\nabla_{\|}h_\alpha\big)^2$ the kernel can be written as a
functional of the height profile. For the same reason as the factors
$\eta_\alpha(\zeta,\ybf_\|)$ could be omitted above, we can neglect
the normalization factor $(-1)^\alpha/\sqrt{g_\alpha}$ of the normal
vector. Thus we obtain for the differential operator
\begin{equation}\label{eq:big_epsilon_matrix}
\big[\hat{\nbf}^\alpha_k\,\hat\Lsc^{k\alpha}\big](\zeta,\ybf_{\|})\:=\:
\begin{pmatrix}
-ih_{\alpha,2}\pd_3-i\pd_2 & 
0 & -\zeta & -h_{\alpha,2}\zeta\\[5pt]
ih_{\alpha,1}\pd_3+i\pd_1 & 
\zeta & 0 & h_{\alpha,1}\zeta\\[5pt]
-ih_{\alpha,1}\pd_2+ih_{\alpha,2}\,\pd_1 & 
h_{\alpha,2}\zeta & -h_{\alpha,1}\zeta & 0
\end{pmatrix}
\:\equiv\:\tilde{\Ombf}^\alpha.
\end{equation}
We observe that the third row of the matrix in
(\ref{eq:big_epsilon_matrix}) is linearly dependent since
$h_{\alpha,1}\tilde{\Ombf}^\alpha_{1\mu}+
h_{\alpha,2}\tilde{\Ombf}^\alpha_{2\mu}=\tilde{\Ombf}^\alpha_{3\mu}$.
Therefore, as in the Lifshitz theory limit discussed earlier, the
matrix has to be reduced to its first two rows. The linear dependence
of rows reflects the fact that for ideal metals there are, in fact,
only two independent boundary conditions for each surface. As
mentioned earlier, for general deformations a reduction to TE and TM
modes as it appears in Lifshitz theory is not possible. However, for
ideal metals the boundary conditions can be simply written as
\begin{equation}
\left[ \hat\nbf_\alpha(\ybf) \times \Ebf(\zeta,\ybf) 
\right]_{\ybf \in S_\alpha}=0.
\end{equation}
This boundary condition requires the two tangential components of the
electric field at the surface to vanish locally. Saying it
differently, the limit of infinite conductivity converts the three
originally non-local boundary condition into two local conditions.
The final result for the matrix kernel is now given by a
$4\times4$--matrix with $j$, $l=1$, $2$,
\begin{equation}\label{eq:big_epsilon_kernel2}
\tilde\Mcal^{\alpha\beta,jl}\big(\zeta,\ybf_{\|};\zeta',\ybf_{\|}'\big)\:=\:
2\pi \delta(\zeta-\zeta')
\big[\tilde{\Ombf}^\alpha\!\cdot\tilde{\Ombf}'^{\beta\dagger}\big]_{jl}
\,\Gcal\big(\zeta;\ybf-\ybf'\big)\bigg\vert_
{\begin{smallmatrix}
y_3=h_\alpha(\ybf_{\|})+H_\alpha\\
y_3'=h_\beta(\ybf_{\|}')+H_\beta
\end{smallmatrix}}
\end{equation}
with the deformation dependent differential operator
\begin{equation}\label{eq:big_epsilon_new_kernel}
\big[\tilde{\Ombf}^\alpha\!\cdot\tilde{\Ombf}'^{\beta\dagger}
\big]_{jl}\:=\:(-1)^{j+l}\Big(\zeta^2\big[\delta_{jl}+
h_{\alpha,3-j}h_{\beta,3-l}'\big]+
\big[h_{\alpha,3-j}\pd_3+\pd_{3-j}\big]
\big[h_{\beta,3-l}'\pd'_3+\pd'_{3-l}\big]\Big)
\end{equation}
acting on the vacuum Green function.  Here the prime on $\:h\:$
indicates the dependence on the primed variable $\ybf'_\|$. This
kernel together with the formula of Eq.~(\ref{eq:free-energy-gen})
yields the exact free energy of the interacting surfaces.  Generally
it is not possible to give a closed analytical expression for the
determinant of $\:\tilde\Mcal\:$ of (\ref{eq:big_epsilon_kernel2}).
However, either perturbative \cite{Emig+01,Emig+03} or numerical
\cite{emig} techniques can be used to evaluate the free energy and
force from the kernel using Eq.~(\ref{eq:free-energy-gen}) and
Eq.~(\ref{eq:force-gen}).

\section{Discussion and Outlook}
\label{sec:discussion}

We derived from a path-integral quantization of the electromagnetic
gauge field with non-local boundary conditions an effective action for
the molecular forces between dielectric media with deformed surfaces.
From this effective action we derived the Lifshitz formula for flat
surfaces and an explicit expression for the matrix kernel which
determines the interaction between ideal metals with arbitrary
deformations. We believe that our approach will be a useful starting
point for the calculation of correlations between modifications of the
ideal Casimir force of Eq.~(\ref{eq:original-force}) due to material
properties (finite conductivity) and geometry.  Our theory is
certainly not a final answer to the problem of correlation effects.
Rather it is intended for a generalization of the Lifshitz formula to
deformed surfaces.  As for the Lifshitz formula, for an explicit
calculation of the force one, of course, has to specify a suitable
dielectric function and, in addition, a height profile describing the
geometry. An precise result for the force can presumably only be
obtained from a numerical evaluation of the effective action. However,
certain limiting cases should be accessible to a perturbative
analysis. For example, it would be interesting to consider the cases
of small and large distances $H$ between the surfaces separately. For
small $H$, the interaction is dominated by high frequencies and is
sensitive to the model from which the frequency dependence of the
dielectric function $\epsilon(\omega)$ is obtained. The correct choice
of $\epsilon(\omega)$ for real metals is currently being discussed but
this is not the matter of the present discussion. For large $H$ a
useful approximation is to assume $\epsilon$ to be constant. In a
large $\epsilon$ expansion a strong geometry dependence and
non-additivity of Casimir forces between ideal metals should be found.
In the opposite limit of rarefied media the non-additivity effects and
thus the sensitivity to geometry should decrease if $\epsilon$
approaches one. Another point of recent dispute is the choice of
boundary conditions for real metals \cite{Geyer+03}. Although this is
not the subject of the present paper, we note that our path integral
approach is sufficiently general to include any type of boundary
conditions as, e.g., the so-called impedance boundary condition
\cite{Bezerra+02a,Bezerra+02b}.

\appendix

\section{Scalar field theory for flat surfaces}
\label{sec:scalar-field}

The scope of application of scalar field theories is more restricted
than that of the gauge field formalism, but calculations become
simpler in many cases. An interesting example is the case of flat
surfaces of dielectric media. In Section \ref{sec:lt} we re-derived
the Lifshitz theory for dielectric media. There we argued that the
field can be decomposed into TE and TM modes each of which have to
obey only one boundary condition on each surface, leading to a reduced
matrix kernel.  In this Appendix we show explicitly that the Lifshitz
theory can be obtained directly from a path integral quantization of a
scalar field which fulfills a suitable boundary condition. Therefore,
no gauge fixing procedure is needed in this special situation.

In order to decompose the electromagnetic field into transversal
electric (TE) and magnetic (TM) modes, we have to specify a preferred
spatial direction. Due to the rotational symmetry in the lateral plane
of the flat surface geometry, we can choose without any restrictions
the $y_2$ direction. Actually, it turns out that it is useful to adapt
the choice of the direction to the lateral momentum $\kbf_\|$ of the
field mode. Note that we can do this since the modes for different
$\kbf_\|$ are decoupled for flat surfaces. Therefore, in the following
we will choose for a given $\kbf_\|$ the lateral coordinates so that
$k_2=0$, and $y_2$ defines the longitudinal direction. Following the
parameterization of the TE and TM modes for waveguides \cite{jackson},
the longitudinal components of the electric (TM modes) and magnetic
(TE modes) field define a (real-valued) scalar field $\Phi$,
\begin{equation}
\label{eq:TM-TE-def}
\begin{array}{lll}
E_2=\Phi\quad&\quad B_2=0
\quad&\text{for TM modes}\\[5pt]
B_2=\Phi\quad&\quad E_2=0
\quad&\text{for TE modes}.
\end{array}\end{equation}
The transversal components $E_j\equiv E_j(\zeta,\kbf_\|;z)$ and
$B_j\equiv B_j(\zeta,\kbf_\|;z)$ of the electromagnetic field are then
given by
\begin{equation}\begin{array}{ll}\label{eq:TM}
E_1\:=\:\frac{k_1k_2}{\zeta^2+k_2^2}\:\Phi\quad & B_1\:=\:
\frac{\zeta}{\zeta^2+k_2^2}\:\pd_z\Phi\\[8pt]
E_3\:=\:\frac{-ik_2}{\zeta^2+k_2^2}\:\pd_z\Phi\quad & B_3\:=\:
\frac{-i\zeta k_1}{\zeta^2+k_2^2}\:\Phi
\end{array}\end{equation}
for TM modes, and
\begin{equation}\begin{array}{ll}\label{eq:TE}
E_1\:=\:\frac{-\zeta}{\zeta^2+k_2^2}\,\pd_z\Phi\quad & B_1\:=\:
\frac{k_1k_2}{\zeta^2+k_2^2}\:\Phi\\[8pt]
E_3\:=\:\frac{i\zeta k_1}{\zeta^2+k_2^2}\,\Phi\quad & B_3\:=\:
\frac{-ik_2}{\zeta^2+k_2^2}\:\pd_z\Phi
\end{array}\end{equation}
for TE modes, where we performed already a Wick rotation to imaginary
frequency, $\omega \to i\zeta$. Using Maxwell equations it can be
shown that the dynamics of the scalar field $\Phi$ are governed by the
usual wave equation, corresponding to the Euclidean action
\begin{equation}
  \label{eq:scalar-action}
  S_\text{E}[\Phi]=\frac{1}{2}\int_X (\nabla \Phi)^2
\end{equation}
with the partition function 
\begin{equation}
\label{eq:scalar-Z-0}
\Zcal_0\:=\:\int\!\Dsc \Phi \:
e^{-S_\text{E}[\Phi]}.
\end{equation}
In this section we consider only the zero temperature case. Finite
temperatures can be treated in analogy to the gauge field approach in
section \ref{sec:general-result} by introducing Matsubara frequencies.
The boundary condition for $\Phi$ can be derived from the general
condition we found in section \ref{sec:bcs}.  From Eq.~(\ref{eq:EX2})
we obtain, using the Fourier representation of the material Green
function $\Gcal^\epsilon$, in Euclidean space the conditions
\begin{eqnarray}
\label{eq:scalar_BC}
-\zeta \epsilon_\alpha(i\zeta) E_2-
ik_1B_3 - (-1)^\alpha p_\alpha B_1 & = & 0
\\[5pt]
\zeta \epsilon_\alpha(i\zeta) E_1-
ik_2B_3 - (-1)^\alpha p_\alpha B_2 & = & 0
\\[5pt]
i\kbf_\|\cdot\Bbf_{\|} - (-1)^\alpha p_\alpha B_3 & = & 0
\end{eqnarray}
for the flat surface $S_\alpha$ for $\alpha=1$, $2$, and with
$p_\alpha=\sqrt{\epsilon_\alpha(i\zeta)\zeta^2+\kbf_\|^2}$. Now we
make use of the fact that we can constrict the analysis to the case
$k_2=0$ by a suitable choice of the lateral coordinates.  After
inserting the electromagnetic field components as given by
Eqs.~(\ref{eq:TM-TE-def}), (\ref{eq:TM}) and (\ref{eq:TE}), the above
conditions collapse to a single boundary condition for the scalar
field $\Phi$. Depending on the type of mode, we obtain the condition
\begin{equation}\label{eq:scalar_LBC}
\Big[1-\Gamma_{\alpha}\,\pd_{n_{\alpha}}\!\Big]\Phi|_{z=H_\alpha}
\:=\:0
\end{equation}
with
\begin{equation}
  \label{eq:def-Gamma}
  \Gamma_\alpha=\frac{1}{p_\alpha} \quad\text{for TM modes},\quad
  \Gamma_\alpha=\frac{\epsilon_\alpha}{p_\alpha} \quad\text{for TE modes},
\end{equation}
where $\pd_{n_{\alpha}}=(-1)^{\alpha-1}\pd_z$ denotes the normal
derivatives of the surfaces.  Due to the $\kbf_\|$--dependence, the
condition (\ref{eq:scalar_LBC}) is non-local in position space.  In
the limit of ideal metals, $\epsilon_\alpha\ra\infty$, the above
condition reduces to the well known Dirichlet and Neumann boundary
conditions for TM and TE modes, respectively.  We note that
$\Gamma_\alpha$ is real, since the same is valid for $\epsilon_\alpha$
on the imaginary frequency axis.  Therefore we need to consider only a
real-valued field $\Phi$.  The restricted partition function for this
field  reads for both types of modes
\begin{equation}
\Zcal \:=\:\Zcal^{-1}_0\int\!\Dsc \Phi \:
e^{-S_E\{\Phi\}}\prod_\alpha\prod_{\zeta, \kbf_\|}
\delta\left[(1-\Gamma_{\alpha}\,\pd_{n_{\alpha}})
\Phi|_{z=H_\alpha} \right],
\end{equation}
where we implemented the boundary constraints again by
delta-functions. Now we proceed in analogy to the treatment of the
gauge field path integral in section \ref{sec:general-result}. We
introduce two auxiliary field $\psi_\alpha$, one for each surface, in
order to replace the delta-function. After integrating out $\Phi$, we
obtain for the partition function
\begin{equation}
  \label{eq:scalar-Z-eff}
  \Zcal = \int \Dsc \psi_\alpha \, e^{-S_\text{eff}[\psi_\alpha]}
\end{equation}
with the effective action
\begin{equation}
  \label{eq:scalar-S-eff}
  S_\text{eff}[\psi_\alpha]=\frac{1}{2}
\int_{\zeta,\kbf_\|} \int_{\zeta',\kbf'_\|}
\psi_\alpha(\zeta,\kbf_\|) 
\Mcal^{\alpha\beta}(\zeta,\kbf_\|;\zeta',\kbf'_\|) \psi_\beta(\zeta',\kbf'_\|)
\end{equation}
and the $2\times 2$ matrix kernel
\begin{equation}
\label{eq:scalar-kernel}
\Mcal^{\alpha\beta}\big(\zeta,\kbf_\|;\zeta',\kbf_\|'\big)\:=\:
(2\pi)^3\,\delta(\zeta+\zeta')\,\delta\big(\kbf_\|+\kbf_\|'\big)
\:\frac{1}{2p}
\twomatrix{1-p^2\Gamma_1^2}
{(1-p\Gamma_1)(1-p\Gamma_2)\,e^{-pH}}
{(1-p\Gamma_1)(1-p\Gamma_2)\,e^{-pH}}
{1-p^2\Gamma_2^2},
\end{equation}
with $k=|\kbf_\| |$, $p=\sqrt{\zeta^2+k^2}$ and a
summation over $\alpha$ and $\beta$ is implicit. From this kernel, the
Casimir energy can be quite easily calculated. Using
Eq.~(\ref{eq:free-energy-gen}) in the zero temperature limit, we
obtain for the Casimir energy per unit surface area
\begin{equation}\label{eq:scalar_lifshitzenergy}
\Fcal(H)\:=\:\frac{1}{2}\int_\zeta \!
\int_{\kbf_\|}\!\ln |\Mcal \Mcal_\infty^{-1}|
\:=\:\int_{0}^{\infty} \frac{d\zeta}{2\pi}
\int_{0}^{\infty}\!\frac{k\,dk}{2\pi}\:\ln\bigg[\,
1-e^{-2pH}\frac{1-p\Gamma_{1}}{1+p\Gamma_{1}}\,
\frac{1-p\Gamma_{2}}{1+p\Gamma_{2}}\,\bigg],
\end{equation}
where $|\ldots |$ denotes the determinant of the $2 \times 2$ matrix
at fixed $\zeta$ and $\kbf_\|$. The matrix $\Mcal_\infty$ is diagonal
since for $H\to\infty$ the off-diagonal elements in
Eq.~(\ref{eq:scalar-kernel}) vanish. Therefore, the determinant of the
matrix product can be easily obtained, yielding the last expression in
Eq. (\ref{eq:scalar_lifshitzenergy}). If we substitute $\Gamma_\alpha$
according to Eq.~(\ref{eq:def-Gamma}) in
Eq.~(\ref{eq:scalar_lifshitzenergy}) by $1/p_\alpha$ or by
$\epsilon_\alpha/p_\alpha$ we obtain the contribution of the TM or TE
modes, respectively, to the energy. It is easily seen that the sum of
the energies from both modes reproduces the Lifshitz theory result we
obtained before in section \ref{sec:lt} from the gauge field approach,
cf. Eq.~(\ref{eq:energie}).

\acknowledgments

This work was supported by the Deutsche Forschungsgemeinschaft through
the Emmy Noether grant No. EM70/2-2.

\end{document}

%% file: makros2.tex
\newcommand{\Fcal}{\mathcal F}
\newcommand{\Gcal}{\mathcal G}

\newcommand{\Lcal}{\mathcal L}
\newcommand{\Mcal}{\mathcal M}

\newcommand{\Zcal}{\mathcal Z}

\newcommand{\Dsc}{\mathscr D}

\newcommand{\Lsc}{\mathscr L}

\newcommand{\Bbf}{\mathbf {B}}

\newcommand{\Ebf}{\mathbf {E}}

\newcommand{\kbf}{\mathbf {k}}

\newcommand{\nbf}{\mathbf {n}}

\newcommand{\qbf}{\mathbf {q}}

\newcommand{\xbf}{\mathbf {x}}

\newcommand{\ybf}{\mathbf {y}}

\newcommand{\Ombf}{\mathbf {\Omega}}

\newcommand{\ra}{\rightarrow}

\newcommand{\abs}[1]{| #1 |}

\newcommand{\til}[1]{\tilde{#1}}

\newcommand{\pd}{\partial}

\newcommand{\twomatrix}[4]{\begin{pmatrix} #1 & #2 \\[5pt] #3 & #4 \end{pmatrix}}